\documentclass[a4paper]{jpconf}

\usepackage{ifthen}      
\usepackage{longtable}   
\usepackage{mathtools}
\usepackage{rotating}    
\usepackage{graphpap}    
\usepackage{multirow}    
\usepackage{mathrsfs}    
\usepackage{pifont}      
\usepackage{afterpage}   
\usepackage{lscape}
\usepackage{microtype}
\usepackage{lineno}      
\usepackage{xspace}      
\usepackage{caption}     
\usepackage{graphicx}    
\usepackage{color}
\usepackage{colortbl}
\graphicspath{{./figs/}} 
\usepackage{amsmath}     
\usepackage{amssymb}
\usepackage{amsfonts}
\usepackage{upgreek}     
\usepackage{hyperref}    
\usepackage[all]{hypcap} 

\def\lhcb     {\mbox{LHCb}\xspace}
\def\cms      {\mbox{CMS}\xspace}

\def\cdf      {\mbox{CDF}\xspace}
\def\dzero    {\mbox{D0}\xspace}

\def\CP      {{\ensuremath{C\!P}}\xspace}

\newcommand{\tev}       {\ensuremath{\mathrm{\,Te\kern -0.1em V}}\xspace}

\newcommand{\gev}       {\ensuremath{\mathrm{\,Ge\kern -0.1em V}}\xspace}
\newcommand{\mev}       {\ensuremath{\mathrm{\,Me\kern -0.1em V}}\xspace}
\newcommand{\mevcc}     {\ensuremath{{\mathrm{\,Me\kern -0.1em V\!/}c^2}}\xspace}
\newcommand{\gevc}      {\ensuremath{{\mathrm{\,Ge\kern -0.1em V\!/}c}}\xspace}
\newcommand{\gevcc}     {\ensuremath{{\mathrm{\,Ge\kern -0.1em V\!/}c^2}}\xspace}

\def\PQ         {\ensuremath{\mathrm{Q}}\xspace}
\def\Quark     {{\ensuremath{\PQ}}\xspace}
\def\Quarkbar  {{\ensuremath{\overline \Quark}}\xspace}
\def\QQbar     {{\ensuremath{\Quark\Quarkbar}}\xspace}

\def\Pchi       {\ensuremath{\upchi}\xspace}

\def\Peta       {\ensuremath{\upeta}\xspace}

\def\Pb         {\ensuremath{\mathrm{b}}\xspace}
\def\bquark    {{\ensuremath{\Pb}}\xspace}
\def\bquarkbar {{\ensuremath{\overline \bquark}}\xspace}
\def\bbbar     {{\ensuremath{\bquark\bquarkbar}}\xspace}

\def\Pmu                {\ensuremath{\upmu}\xspace}

\def\mup               {{\ensuremath{\Pmu^+}}\xspace}
\def\mumu              {{\ensuremath{\Pmu^+\Pmu^-}}\xspace}

\def\deriv              {\ensuremath{\mathrm{d}}}
\def\sqs                {\ensuremath{\protect\sqrt{s}}\xspace}
\def\invfb              {\ensuremath{\mbox{\,fb}^{-1}}\xspace}
\newcommand{\chisq}     {\ensuremath{\chi^2}\xspace}

\def\Pp                 {\ensuremath{\mathrm{p}}\xspace}
\def\proton            {{\ensuremath{\Pp}}\xspace}
\def\antiproton        {{\ensuremath{\overline \proton}}\xspace}
\def\PJ                 {\ensuremath{\mathrm{J}}\xspace}
\def\Ppsi               {\ensuremath{\uppsi}\xspace}

\def\jpsi  {{\ensuremath{{\PJ\mskip -3mu/\mskip -2mu\Ppsi\mskip 2mu}}}\xspace}
\def\psitwos  {{\ensuremath{\Ppsi{(2S)}}}\xspace}

\newcommand{\decay}[2]{\ensuremath{#1\!\to #2}\xspace}
\def\Pmu              {\ensuremath{\upmu}\xspace}
\def\mumu            {{\ensuremath{\Pmu^+\Pmu^-}}\xspace}

\def\PUpsilon           {\ensuremath{\Upsilon}\xspace}
\newcommand{\upsmm}     {\decay{\PUpsilon}{\mumu}}
\newcommand{\ups}       {\PUpsilon}
\newcommand{\upsns}     {\ensuremath{\PUpsilon(\mathrm{nS})}\xspace}
\newcommand{\ones}      {\ensuremath{\PUpsilon(1\mathrm{S})}\xspace}
\newcommand{\twos}      {\ensuremath{\PUpsilon(2\mathrm{S})}\xspace}
\newcommand{\threes}    {\ensuremath{\PUpsilon(3\mathrm{S})}\xspace}

\def\Quark             {{\ensuremath{\PQ}}\xspace}
\def\Quarkbar          {{\ensuremath{\overline \Quark}}\xspace}
\def\QQbar             {{\ensuremath{\Quark\Quarkbar}}\xspace}
\def\ptot               {\mbox{$p$}\xspace}
\newcommand{\bit}{\begin{itemize}}
\newcommand{\bce}{\begin{center}}
\newcommand{\eit}{\end{itemize}}
\newcommand{\ece}{\end{center}}

\newcommand{\pT}     {\ensuremath{p_{\mathrm{T}}}\xspace}

\def\sPlot{\mbox{\em sPlot}}
\def\sWeights{\mbox{\em sWeights}}

\newcommand{\pty}  {\ensuremath{p^{\ups}_{\mathrm{T}}}\xspace}
\newcommand{\xfy}  {\ensuremath{x^{\ups}_{\mathrm{F}}}\xspace}
\newcommand{\yy}   {\ensuremath{y^{\ups}}\xspace}

\newcommand{\MYCIRCLE} {\ensuremath{\textrm{\ding{108}}}}

\newcommand{\MYDIAMOND}{\ensuremath{\textrm{\ding{117}}}}

\begin{document}
\title{\upsns polarizations in \proton\proton
  collisions at~\mbox{\sqs=\,7} and~\mbox{8\tev} by the \lhcb collaboration}

\author{Alexander Artamonov on behalf of the \lhcb Collaboration}

\address{Institute for High Energy Physics, National Research Center
  Kurchatov Institute, Protvino, Moscow region, 142281 Russia}

\ead{Alexander.Artamonov@ihep.ru}

\begin{abstract}
A polarization measurement carried out for the~\ones, \twos and
\threes~mesons produced in pp collisions at $\sqs=7$ and~\mbox{8\tev}
is presented.
Data samples used for the polarization measurement were collected by
the \lhcb experiment during the 2011 and 2012 data taking runs with
integrated luminosities of 1 and 2\invfb, respectively.
The measurement has been performed in three polarization frames,
using an angular distribution analysis of the
$\ups\to\mumu$~decays 
in the kinematic region of the \ups transverse momentum
\mbox{$\pty<30~\gevc$} and rapidity \mbox{$2.2<\yy<4.5$}.
No large polarization is observed.
\end{abstract}

\section{Introduction}
It is already forty years since the discovery of the first bottomonium
state, the~\ones meson~\cite{Lederman1977},
but studies of heavy quarkonium production continue
to play an important role in the development of quantum chromodynamics
(QCD)~\cite{Braaten_Russ_2014}.
According to the current theoretical framework,
nonrelativistic QCD 
(NRQCD)~\cite{CaswellLepage1986PL,Bodwin_Braaten_Lepage_1995},
inclusive production of a heavy quarkonium
is viewed as a two-step process.
In the first step, a heavy quark-antiquark pair, \QQbar,
is perturbatively created in a color singlet or color octet state,
and then, in the second step,
the hadronization process non-perturbatively
transforms the \QQbar pair into an observable colorless bound state.
The non-perturbative transitions
are described by long-distance matrix elements
which are conjectured to be independent of production processes
in the first step,
and need to be extracted experimentally.
The most distinct signature of the first NRQCD
calculations~\cite{Cho_Wise_1994,Beneke_Rothstein_1995,Beneke_Kramer_1996,Leibovich_1996}
was the prediction of
transverse polarization for S-wave quarkonium states
(such as the \jpsi, \psitwos and \upsns mesons)
directly produced (i.e. not coming from decays)
at large transverse momentum
in high-energy hadron collisions.
NRQCD calculations for promptly produced heavy quarkonia
(if charmonia then those not coming from b-hadron decays)
are complicated by 
feed\nobreakdash-down (electromagnetic or hadronic transitions)
from higher level states.

Full NLO calculations~\cite{Gong:2013qka}
performed for the \ones, \twos and \threes
mesons (shortly denoted by \ups),
including effects of feed\nobreakdash-down contributions
for the \ones and \twos but not for the \threes,
give very small transverse polarizations for the first two
bottomonium states and a large transverse polarization
for the third state.
These calculations were done before \lhcb obtained that
the fractions of \ups mesons originating from
$\Pchi_{\bquark}$~decays are around
$30-40\,\%$ for high transverse momenta,
$\pty\gtrsim20\gevc$~\cite{LHCb-PAPER-2014-031}.
The \threes was considered as almost
feed\nobreakdash-down free state before.
Although taking into account feed\nobreakdash-down contributions
improves the description of \ups polarization,
there are still problems in describing the
\jpsi polarization~\cite{LHCb_psi_polar}
(even with considering feed\nobreakdash-down)
and the \psitwos polarization~\cite{LHCb_psi2S_polar}
(which includes negligible feed\nobreakdash-down contributions).

The angular distribution of muons from the $\ups\to\mumu$~decay
can be written as~\cite{Oakes1966,LamTung1978,Faccioli_clarification}
\begin{equation}
 \frac{1}{\sigma}\frac{\deriv\sigma}{\deriv\Omega} =
 \frac{3}{4\pi}~\frac{1}{3 + \uplambda_{\theta}}
 \left(1 + \uplambda_{\theta}\cos^2\theta 
         + \uplambda_{\theta\phi}\sin2\theta\cos\phi 
         + \uplambda_{\phi}\sin^2\theta\cos2\phi \right),
\label{eq:MainAngDistrib}
\end{equation}
where the angular quantities $\Omega = \left(\cos\theta, \phi \right)$
describe a direction of \mup in the \ups rest frame
with respect to some specified axes,
$\vec{\uplambda} \equiv
( \uplambda_{\theta}, \uplambda_{\theta\phi}, \uplambda_{\phi} )$
are the angular distribution parameters
directly related to the spin-1 density-matrix
elements~\cite{Oakes1966,Pilkuhn1979,Beneke_Kramer_Vanttinen_1998}.
The parameter $\uplambda_{\theta}$ is a measure of spin-alignment,
and can be expressed as
\mbox{$\uplambda_{\theta} = \left( \sigma_{T} - 2\sigma_{L}
\right)/ \left( \sigma_{T} + 2\sigma_{L} \right)$}, where
$\sigma_{T}$ ($\sigma_{L}$) is the transverse (longitudinal)
component of the cross section.
If the spin-alignment parameter
$\uplambda_{\theta} > 0$ ($\uplambda_{\theta} < 0$),
the $\ups$ meson is called to be transversely (longitudinally)
polarized in a specified frame, while the case
$\uplambda_{\theta} = \uplambda_{\theta\phi} = \uplambda_{\phi} = 0$
means that the $\ups$ meson is unpolarized.
The parameters $\vec{\uplambda}$
depend on a definition of coordinate axes specified in
the $\ups$ rest frame.
The following three coordinate systems are widely used
in polarization analyses:
helicity (HX)~\cite{Jacob_Wick:1959},
Collins-Soper (CS)~\cite{Collins_Soper} and
Gottfried-Jackson (GJ)~\cite{Collins_Soper}.
The frames are specified by different directions of the
spin-quantization axis, $z$~axis, defined in 
the production plane of \ups meson~\cite{Faccioli_clarification}.
In all these frames, the $y$~axis is normal to
the production plane~\cite{Faccioli_clarification,LHCb_YnS_polar_2017},
and the remaining $x$~axis completes a right-handed coordinate system.

Until recently, quarkonium polarization measurements have been
reduced only to studies of the parameter $\uplambda_{\theta}$,
some times in different polarization frames.
As pointed out in~\cite{Faccioli_clarification},
measuring all the three polarization
parameters is important from the theoretical and
experimental points of view. Since from having
the $\vec{\uplambda}$ in one frame, it is possible to transform
them into another~\cite{Faccioli_clarification,Telegdi1986},
and perform some cross checks of results.
In particular, an important cross check is provided by a 
polarization parameter
\mbox{$\tilde{\uplambda} = \left(\uplambda_{\theta} + 
3 \uplambda_{\phi}\right)/
\left(1 - \uplambda_{\phi}\right)$}~\cite{Faccioli_2010_a,Faccioli_2010_b},
which is invariant for all rotations around the $y$ axis,
that is invariant in the HX, CS and GJ frames.
The physical meaning of the parameter $\tilde{\uplambda}$
was first recognized in~\cite{Teryaev:SPIN2005}
(see also~\cite{Teryaev:2011zza,Teryaev:SPIN2011}).

The first full angular distribution analysis of muons
from the $\ups\to\mumu$~decays was performed by
the \cdf collaboration~\cite{CDFpolar2012}
using data of $\proton\antiproton$ collisions
at $\sqrt{s}=1.96$ \tev.
\cdf found that the angular distributions of muons from
all the three \ups states are nearly isotropic
in the central rapidity region \mbox{$|\yy|<0.6$}
and \mbox{$\pty<40~\gevc$}.
This result is consistent with
the previous CDF measurement~\cite{CDFpolar2002},
and inconsistent with the measurement performed by
the \dzero collaboration~\cite{D0polar2008}.
\dzero observed the significant \pty dependent longitudinal
polarization for the \ones mesons produced in $\proton\antiproton$
collisions at $\sqrt{s}=1.96$ \tev, for
\mbox{$|\yy|<1.8$} and \mbox{$\pty<20~\gevc$}.
The next full angular distribution analysis for
the $\ups\to\mumu$~was performed by
the \cms collaboration~\cite{CMSpolar2013}
using $\proton\proton$ collisions data at $\sqrt{s}=7$ \tev,
for the rapidity ranges \mbox{$|\yy|<0.6$}
and \mbox{$0.6<|\yy|<1.2$},
and for \mbox{$10<\pty<50~\gevc$}~\cite{CMSpolar2013}.
CMS found no evidence of large transverse or longitudinal
polarization for any of the three \ups mesons in
the explored kinematic region.
The experimental situation is complicated by
the result of the fixed-target experiment E866~\cite{E866polar2001},
which performed the polarization measurement of
the \ups mesons produced in p-Cu collisions at
$\sqrt{s}=38.8$ \gev in \mbox{$0.0<\xfy<0.6$} and
\mbox{$\pty<4~\gevc$}.
The E866 collaboration found that
the \ones meson is produced weakly polarized,
while the \twos and \threes
mesons are produced with a maximal transverse polarization.
Although different production energies may determine
different dominant contributions in the production processes,
all these results underscore the need for further experimental
study of the \ups polarization.

The \lhcb collaboration performed
the full angular distribution analysis~\cite{LHCb_YnS_polar_2017}
for the \ups mesons produced in $\proton\proton$ collisions at
$\sqrt{s}=7$ and 8\tev in the LHCb setup during the 2011 and
2012 data taking runs with integrated luminosities of
1 and 2\invfb, respectively.
The polarization measurement was done
in the HX, CS and GJ frames
in the \ups kinematic range defined by
\mbox{$\pty<30~\gevc$} and \mbox{$2.2<\yy<4.5$}.

\section{\lhcb detector and selection of \mbox{$\ups\to\mumu$}~decays}
The \lhcb detector is a single-arm forward spectrometer
primarily designed to look for indirect evidence of new physics in
\CP violation and rare decays of charm and beauty hadrons.
The detector covers a unique pseudorapidity range~\mbox{$2<\eta <5$}
corresponding to $\sim 4\%$ of the solid angle, but
$\sim 25\%$ of all produced \bbbar pairs fall into this
geometrical acceptance.
A detailed description of the \lhcb detector
and its performance are given in~\cite{Alves:2008zz,LHCb-DP-2014-002}.

The selection of \ups candidates is similar to those used in
the previous LHCb analyses~\mbox{\cite{LHCb-PAPER-2011-036,
LHCb-PAPER-2013-016,LHCb-PAPER-2013-066,LHCb-PAPER-2015-045}}.
The \ups candidates are formed from pairs of oppositely charged tracks
reconstructed in the tracking system. Each track is required to have
a good reconstruction quality~\cite{LHCb-DP-2013-002}
and to be identified as a muon~\cite{LHCb-DP-2013-001}.
Each muon is then required to have momentum satisfying
\mbox{$1<\pT<25\gevc$}, \mbox{$10<\ptot<400\gevc$}
and pseudorapidity within the region
\mbox{$2.2<\Peta<4.5$}.
The two muons are required to originate from a common vertex
with a good \chisq probability of the vertex fit.
In addition, the consistency of the dimuon vertex with a primary
vertex is ensured via the quality requirement of a global
fit, performed for each dimuon candidate using the primary vertex
position as a constraint~\cite{Hulsbergen:2005pu}.
This global fit requirement also reduces the background
caused by genuine muons coming from decays of long-lived charm
and beauty hadrons.
A large fraction of the combinatorial background is populated
at large values of \mbox{$\left|\cos\theta_{\text{GJ}}\right|$},
where $\theta_{\text{GJ}}$ is the polar angle of $\mup$
in the GJ frame.
To reduce this background, a requirement
\mbox{$\left|\cos\theta_{\text{GJ}}\right|<0.8$}
is applied.
Finally, the mass of the muon pair is required to be in the range
\mbox{$8.8<m_{\mumu}<11.0\,\gevcc$}.

As an example, a dimuon mass distribution of the
\upsmm~candidates finally selected in \mbox{$6<\pty<8\gevc$} and
\mbox{$2.2<\yy<3.0$} for \mbox{$\sqrt{s}=7\,\mathrm{TeV}$}
data set is shown in
Fig.\,1.
The dimuon mass distribution is parametrized
by the sum of three double-sided Crystal Ball
functions~\cite{Skwarnicki:1986xj,LHCb-PAPER-2011-013}
for describing the \ups mesons and an exponential function for
the combinatorial background.
The parametrization is done by an unbinned extended maximum
likelihood fit.
The dimuon mass fit is performed in each $\left(\pty,~\yy\right)$ bin,
and results of the fit are then used by the
\sPlot~analysis~\cite{Pivk_Diberder_2005} for
obtaining so called signal \sWeights:
$w_i^{\ones}$, $w_i^{\twos}$ and $w_i^{\threes}$.
The \sWeights~are assigned to each, $i^{th}$, dimuon candidate
for extracting the appropriate \ups signal component.
The total signal yields
obtained from the dimuon mass fit
in the full explored range of $\pty$ and $\yy$ are
$0.8\times10^{6}$ $(1.8\times10^{6})$ \ones candidates,
$0.2\times10^{6}$ $(0.5\times10^{6})$ \twos candidates and
$0.1\times10^{6}$ $(0.2\times10^{6})$ \threes candidates
in \mbox{$\sqrt{s}=7\,\mathrm{TeV}$}
(\mbox{$\sqrt{s}=8\,\mathrm{TeV}$}) data set.
The average mass resolution of the \ones peak is $42\,\mevcc$.
\begin{figure}[t]
  \setlength{\unitlength}{1mm}
  \centering
  \begin{picture}(150,60)
    \put( 0, 0){
      \includegraphics*[width=75mm,height=60mm]{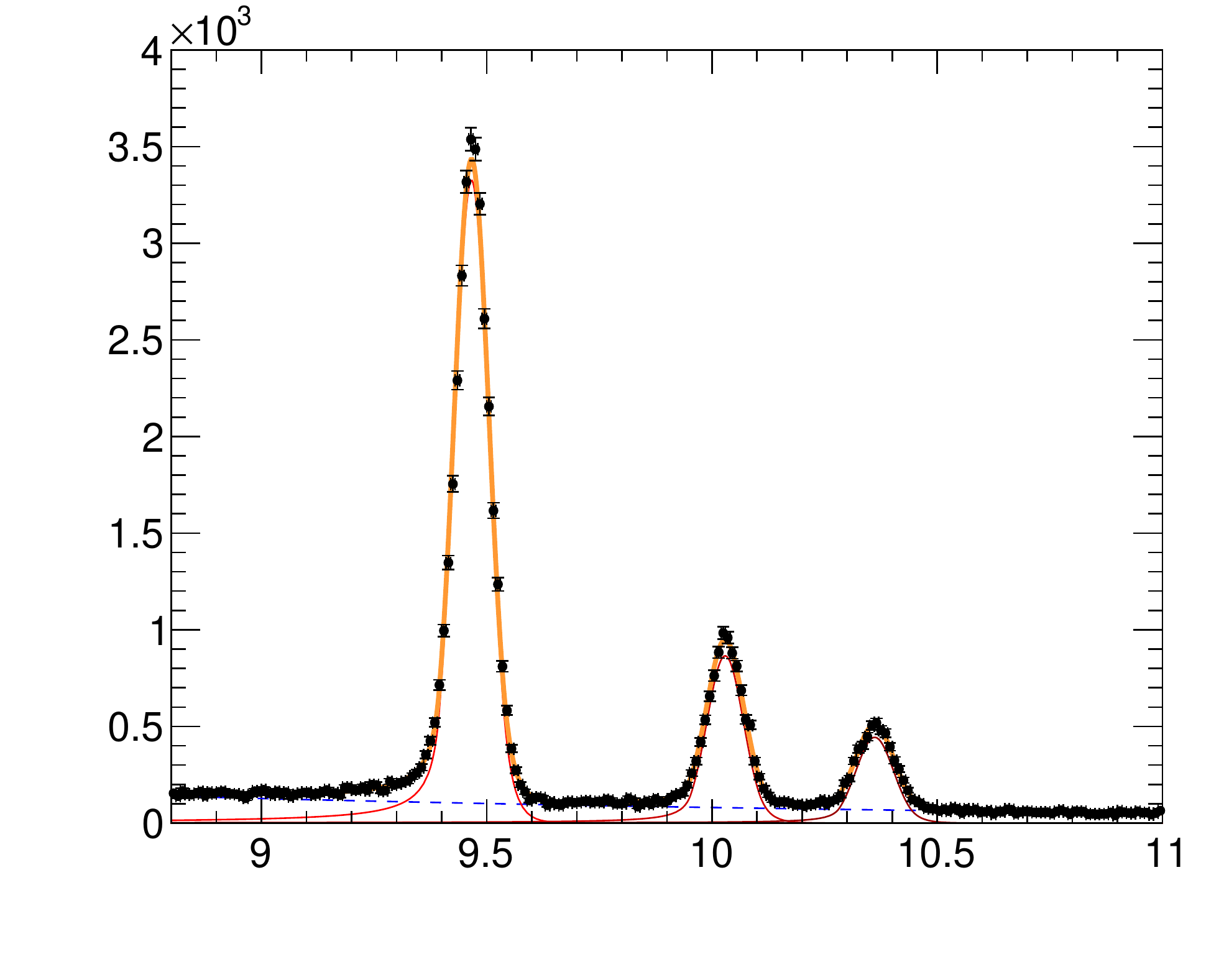}
    }
  \end{picture}
  \put(-68, 33){
    \begin{minipage}{0.42\textwidth}
      \caption {\small
        Dimuon mass distribution
        in the region \mbox{$6<\pty<8\gevc$}, \mbox{$2.2<\yy<3.0$}
        for data obtained at~\mbox{$\sqrt{s}=7\,\mathrm{TeV}$}.
        The thick dark yellow solid curve shows the result of the fit,
        as described in the text.
        The three peaks, shown with thin red solid lines,
        correspond to the \ones, 
        \twos and \threes signals\,(left to right).
        The background component is indicated with a dashed blue line.
      }
    \end{minipage}
  }
  \put(-112, 2) { $m_{\mumu}$}  \put(-94,2) {
    $\left[\!\gevcc\right]$}
  \put(-147,19.0){\begin{sideways}\small Candidates/(10\mevcc)\end{sideways}}
  \put(-113.5, 45.5) { \small { {$\begin{array}{l}\lhcb~\sqs=7\tev
          \\ 6<\pty<8\gevc \\ 2.2<\yy<3.0\end{array}$}}}
  \label{fig:fig01}
\end{figure}
\section{Polarization analysis}
The polarization measurement is performed using an unbinned
maximum likelihood approach~\cite{Xie_2009}
already applied in the \jpsi and \psitwos polarization
analyses~\cite{LHCb_psi_polar},\cite{LHCb_psi2S_polar}.
The polarization parameters are determined from fits to the
two-dimensional
$\left(\cos\theta,\phi\right)$ angular distribution
of \mup from the \mbox{$\ups\to\mumu$}~decay,
described by Eq.\,\ref{eq:MainAngDistrib}.
In each $\left(\pty,~\yy\right)$ bin,
the following logarithm of the likelihood function
is constructed for each \ups state:
\begin{equation}\label{eq:UpsLikelihood_all}
   \log \mathcal{L}(\uplambda_{\theta},\uplambda_{\theta\phi},\uplambda_{\phi})_{\ups} =
   s_{w} \sum^{N_{\mathrm{tot}}}_{i=1} w_i^{\ups} \times
   \log \left[\frac{\mathcal{P}(\cos\theta_{i},\phi_{i} \vert
       \uplambda_{\theta}, \uplambda_{\theta\phi}, \uplambda_{\phi})}
     {\mathcal{N}(\uplambda_{\theta},\uplambda_{\theta\phi}, 
       \uplambda_{\phi})}\right]\;,
\end{equation}
where
$\mathcal{P}(\cos\theta_{i},\phi_{i} \vert \uplambda_{\theta},
\uplambda_{\theta\phi}, \uplambda_{\phi}) \equiv
1+\uplambda_\theta \cos^2 \theta_{i}
+ \uplambda_{\theta \phi}\sin 2\theta_{i} \cos \phi_{i}
+ \uplambda_\phi\sin^2 \theta_{i} \cos 2\phi_{i}$,
$\mathcal{N}(\uplambda_{\theta},\uplambda_{\theta\phi},\uplambda_{\phi})$
is normalization factor determined by Monte Carlo
for each \ups meson,
$w_i^{\ups}$ is one of the \sWeights~for
the $i^{th}$ \ups candidate,
and $N_\mathrm{tot}$ is the total number of all selected
\ups candidates in a considered $\left(\pty,~\yy\right)$ bin.
The constant scale factor
\mbox{$s_{w}=\sum_{i=1}^{N_\mathrm{tot}}w_i^{\ups}/
\sum_{i=1}^{N_\mathrm{tot}}(w_i^{\ups})^2$}
is introduced to take into account correctly the effect of the
\sWeights~on statistical uncertainties of the $\vec{\uplambda}$
parameters obtained after the polarization fit.
The influence of the $s_{w}$ was validated by pseudoexperiments.
The normalization factor
$\mathcal{N}(\uplambda_{\theta},\uplambda_{\theta\phi},\uplambda_{\phi})$
is defined as
\begin{equation}\label{eq:NormFactor_all}
  \mathcal{N}(\uplambda_{\theta},\uplambda_{\theta\phi},\uplambda_{\phi})
  \equiv \int \deriv\Omega~ \mathcal{P}(\cos\theta,\phi \vert
  \uplambda_{\theta},\uplambda_{\theta\phi},\uplambda_{\phi})
  \times \varepsilon(\cos\theta,\phi)
\end{equation}
and is calculated using simulated events. 
In the simulation, where the \ups~mesons are generated unpolarized,
the two\nobreakdash-dimensional $\left(\cos\theta,\phi\right)$
angular distribution of \mup from decays of selected \ups candidates
is proportional to the total efficiency
$\epsilon(\cos\theta,\phi)$, so
$\mathcal{N}(\uplambda_{\theta},\uplambda_{\theta\phi},\uplambda_{\phi})$
is evaluated by summing 
$P(\cos\theta_{j},\phi_{j}\vert\lambda_{\theta}, \lambda_{\theta\phi},
\lambda_{\phi})$ 
over the~selected \ups~candidates in the simulated sample
\begin{equation}
  \mathcal{N}(\uplambda_{\theta},\uplambda_{\theta\phi},\uplambda_{\phi})
  \propto  \sum_{j}
  \epsilon^{\mumu}
  \kappa^{\ups}
  \mathcal{P}(\cos\theta_{j},\phi_{j} \vert
  \uplambda_{\theta},\uplambda_{\theta\phi},\uplambda_{\phi}), \label{eq:NormFactor_4}
\end{equation}
where $\epsilon^{\mumu}$ is a muon identification efficiency
measured directly from data using a large sample of low-background
\mbox{$\jpsi\to\mumu$}~events
(no muon identification requirement was applied when selecting
the \ups candidates in the simulated samples);
$\kappa^{\ups}$ is a correction factor for MC, obtained
using data-driven techniques to account for small differences between
data and simulation in a tracking efficiency of
muons~\mbox{\cite{LHCb-DP-2013-002,LHCb-DP-2013-001}}
and in the \pty and \yy spectra~\cite{Sjostrand:2006za,LHCb-PROC-2010-056}.

\section{Results and conclusions}

Different sources of systematic uncertainty have been considered
when determining the polarization parameters,
namely systematic uncertainty
related to: a) the \ups signal extraction procedure;
b) the muon identification efficiency;
c) the track reconstruction efficiency;
d) a possible small difference in the trigger
efficiency between data and simulation;
e) correction factors for the muon identification efficiency;
f) the finite size of the simulated samples.
All these sources have been studied for the polarization
parameters $\uplambda_{\theta}$, $\uplambda_{\theta\phi}$,
$\uplambda_{\phi}$ and $\tilde{\uplambda}$ in the HX, CS and GJ
frames for each $\left(\pty,~\yy\right)$ bin.
It was found that the most dominant systematic uncertainty
is related to the finite size of MC samples,
varying between $30\%$ and $70\%$ of the statistical uncertainty.
The total systematic uncertainty for each polarization
parameter is calculated as the quadratic sum of systematic
uncertainties coming from all the considered sources,
assuming no correlation between them.
For some high-\pty bins the systematic
and statistical uncertainties are comparable.

All the \ups polarization results obtained by the \lhcb
collaboration for data collected at $\sqrt{s}=7\,\,{\mathrm{TeV}}$
and $8\,{\mathrm{TeV}}$ are given in~\cite{LHCb_YnS_polar_2017}.
Here we outline the main features of the results.
The values of the parameter $\uplambda_{\theta}$
obtained for the \ups mesons
do not show any significant transverse or longitudinal
polarization in all frames
over the considered kinematic region.
The values of the parameters $\uplambda_{\theta\phi}$ and
$\uplambda_{\phi}$ are small
in all frames over all $\left(\pty,~\yy\right)$ bins.
All the three polarization parameters do not manifest
a distinct dependence on the \yy.
The values of the frame invariant parameter
$\tilde{\uplambda}$ measured in the HX, CS and GJ
frames are consistent with each other.
Moreover, all values of the $\tilde{\uplambda}$
are close to zero in all phase-space bins.
In the considered phase space domain,
the \ups polarization results corresponding to
$\sqrt{s}=7$ and 8\tev are in good agreement
with each other.

\begin{figure}[htb]
  \setlength{\unitlength}{1mm}
  \centering
  \begin{picture}(155,120)
    %
    \put(  0,  60){
      \includegraphics*[width=75mm,height=60mm,%
      ]{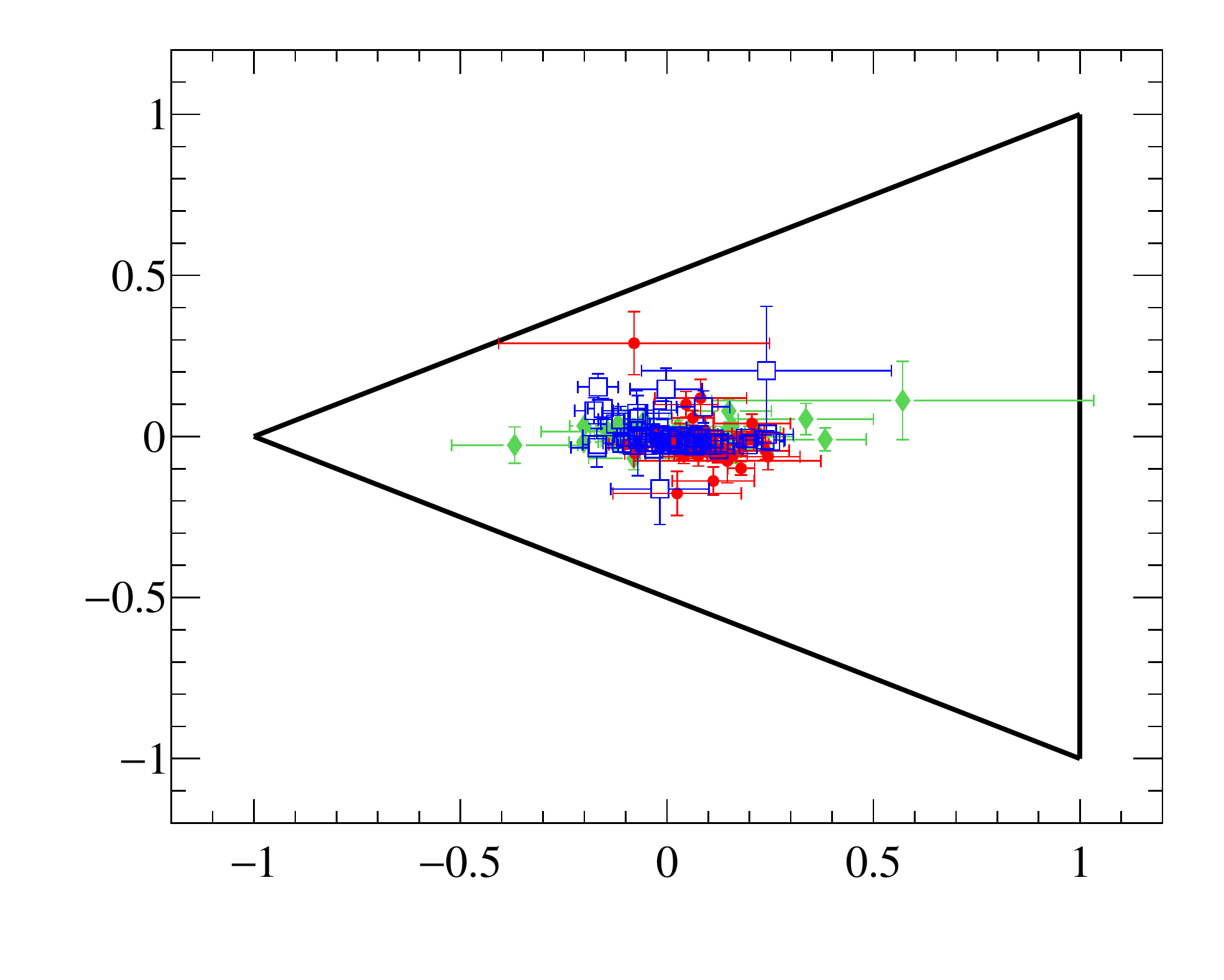}
    }
    \put( 75,  60){
      \includegraphics*[width=75mm,height=60mm,%
      ]{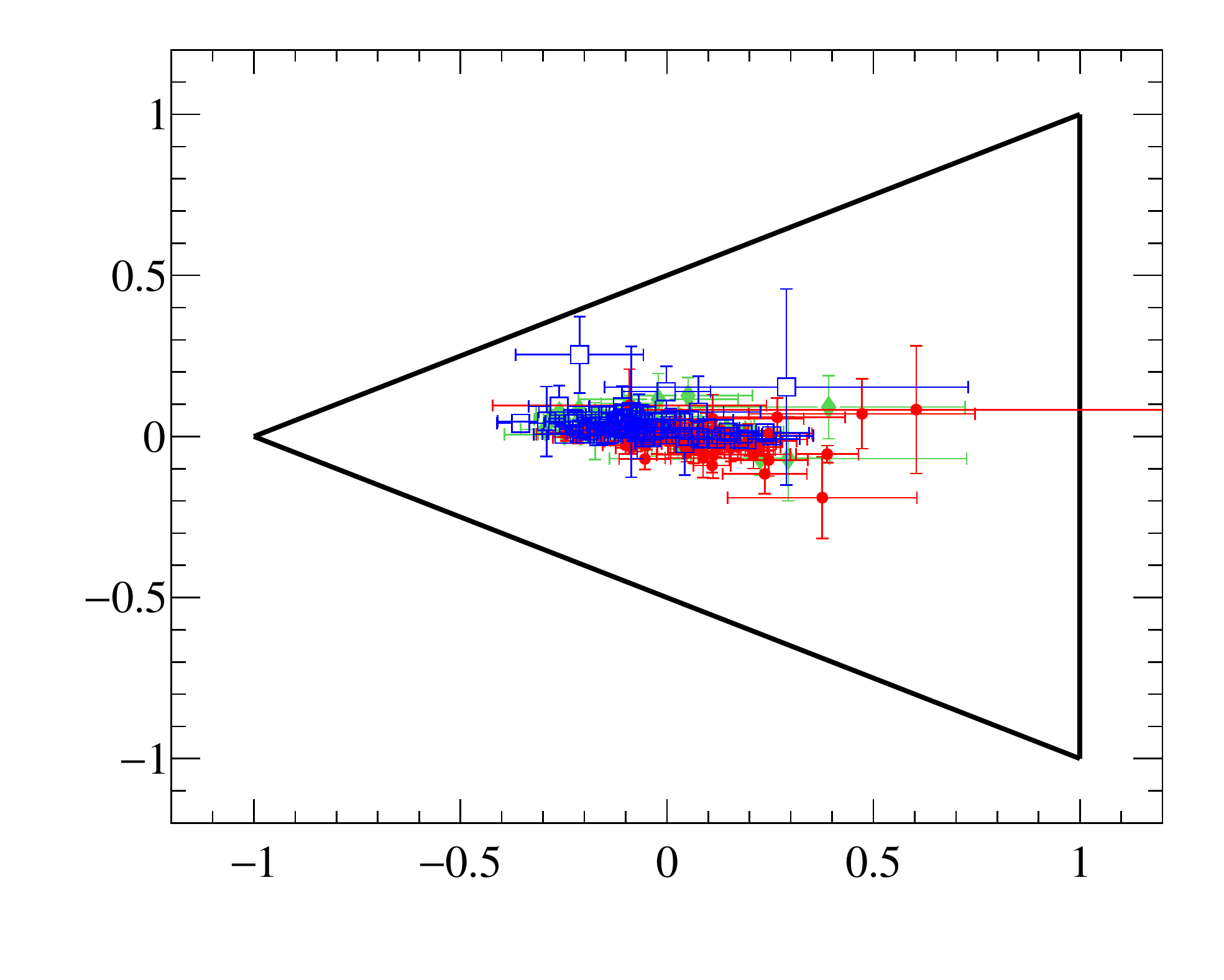}
    }
    \put(  0,   0){
      \includegraphics*[width=75mm,height=60mm,%
      ]{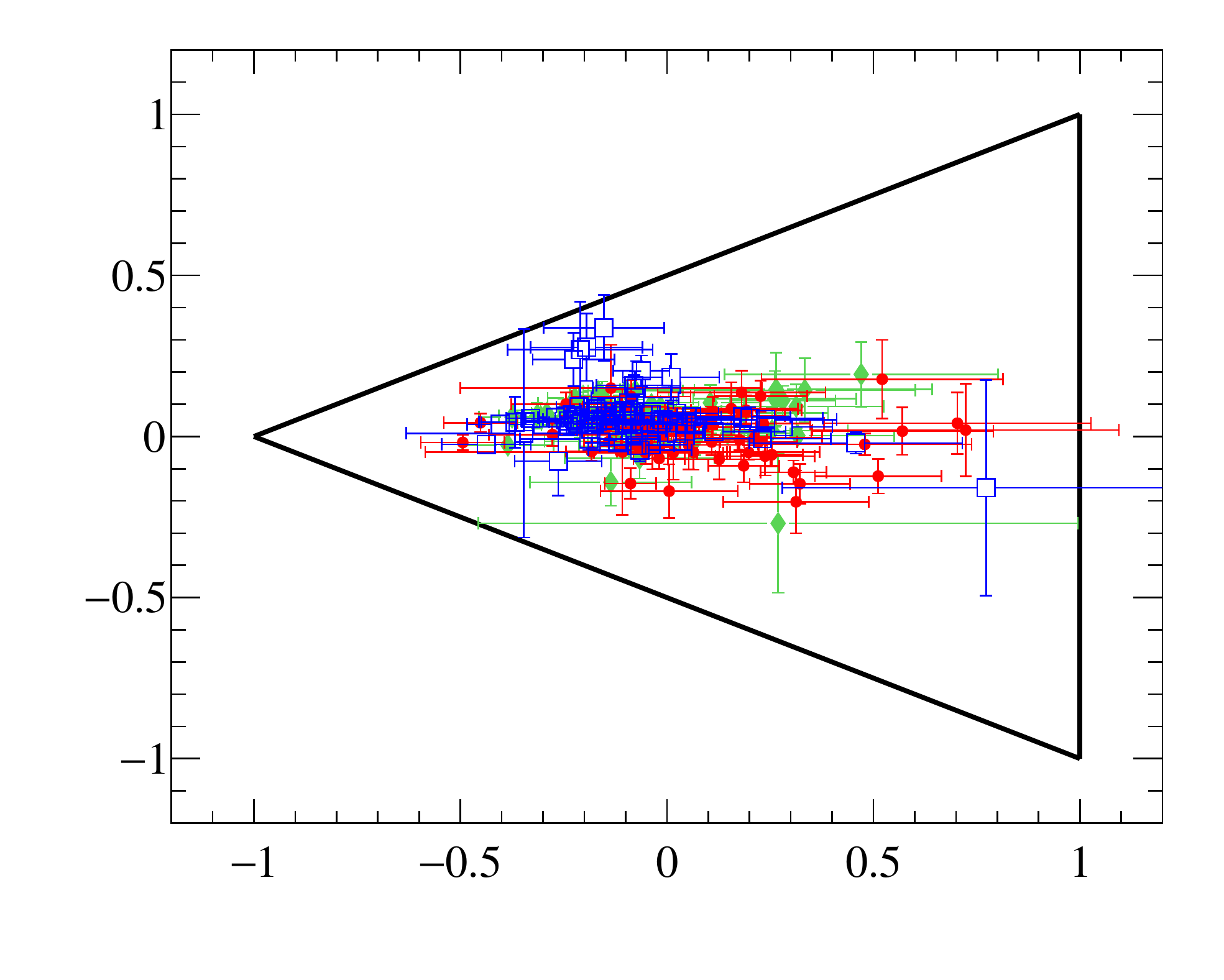}
    }
    \put (13 ,78) { \small $\begin{array}{cl}
        {\color{red}\MYCIRCLE}                 & \mathrm{HX} \\
        {\color{blue}\square}                & \mathrm{CS} \\
        {\color[rgb]{0,0.6,0}\MYDIAMOND}  & \mathrm{GJ} 
      \end{array}$}    
    \put (  0,  95)
         { \begin{sideways}$\uplambda_{\phi}$\end{sideways}}
    \put ( 40,  60) { $\uplambda_{\theta}$}
    \put ( 75,  95)
         { \begin{sideways}$\uplambda_{\phi}$\end{sideways}}
    \put (115,  60) { $\uplambda_{\theta}$}
    \put (  0,  35)
         { \begin{sideways}$\uplambda_{\phi}$\end{sideways}}
    \put ( 40,   0) { $\uplambda_{\theta}$}
    \put (13,108) { a)\small$\begin{array}{l} \lhcb\,\sqrt{s}=7,8\tev
        \\ \ones   \end{array}$}
    \put (88,108) { b)\small$\begin{array}{l} \lhcb\,\sqrt{s}=7,8\tev
        \\ \twos   \end{array}$}
    \put (13, 48) { c)\small$\begin{array}{l} \lhcb\,\sqrt{s}=7,8\tev
        \\ \threes \end{array}$}
  \end{picture}
  \caption { \small
    The measured values of
    $\uplambda_{\theta},\uplambda_{\phi}$~for 
    a)~$\PUpsilon\mathrm{(1S)}$,
    b)~$\PUpsilon\mathrm{(2S)}$ and 
    c)~$\PUpsilon\mathrm{(3S)}$~mesons.
    The~red solid circles, blue open squares and green solid diamonds
    correspond to 
    the~helicity\,(HX), Collins-Soper\,(CS)
    and Gottfried-Jackson\,(GJ) frames, respectively.
    The~thick black lines show
    the~regions allowed by 
    the~positivity constraints.
  }
  \label{fig:positivity}
\end{figure}

The $\vec{\uplambda}$ parameters have been checked for
positivity constraints imposed on the spin-1
density-matrix~\cite{Faccioli_clarification,LHCb_YnS_polar_2017,Palestini:2010xu,Teryaev:2011zza,Teryaev:SPIN2011}.
The values of the $\vec{\uplambda}$ satisfy all the six positivity
constraints in all frames over all phase-space bins.
In particular,
Fig.~\ref{fig:positivity} shows regions allowed by the positivity
constraints together with the parameters $\uplambda_{\theta}$ and
$\uplambda_{\phi}$ measured in all $\left(\pty,~\yy\right)$~bins,
for data collected at~\mbox{$\sqs=7$} and~$8\tev$.
Further, since the $z$ axes of the HX, CS and GJ frames
coincide in the limit
\mbox{$\pty\to0$}~\cite{Faccioli_clarification,Telegdi1986},
we checked this constraint
and found that all values of the $\vec{\uplambda}$ are
very similar for low-\pty bins in all frames.
It was also found that the parameters $\uplambda_{\theta\phi}$
and $\uplambda_{\phi}$ are very close to zero in the limit
\mbox{$\pty\to0$}, in accordance with kinematic constraints
pointed out in~\cite{LamTung1978}.

\begin{figure}[htb]
  \setlength{\unitlength}{1mm}
  \centering
  \begin{picture}(155,120)
    %
    \put(  0,  0){
      \includegraphics*[width=150mm,height=120mm,%
      ]{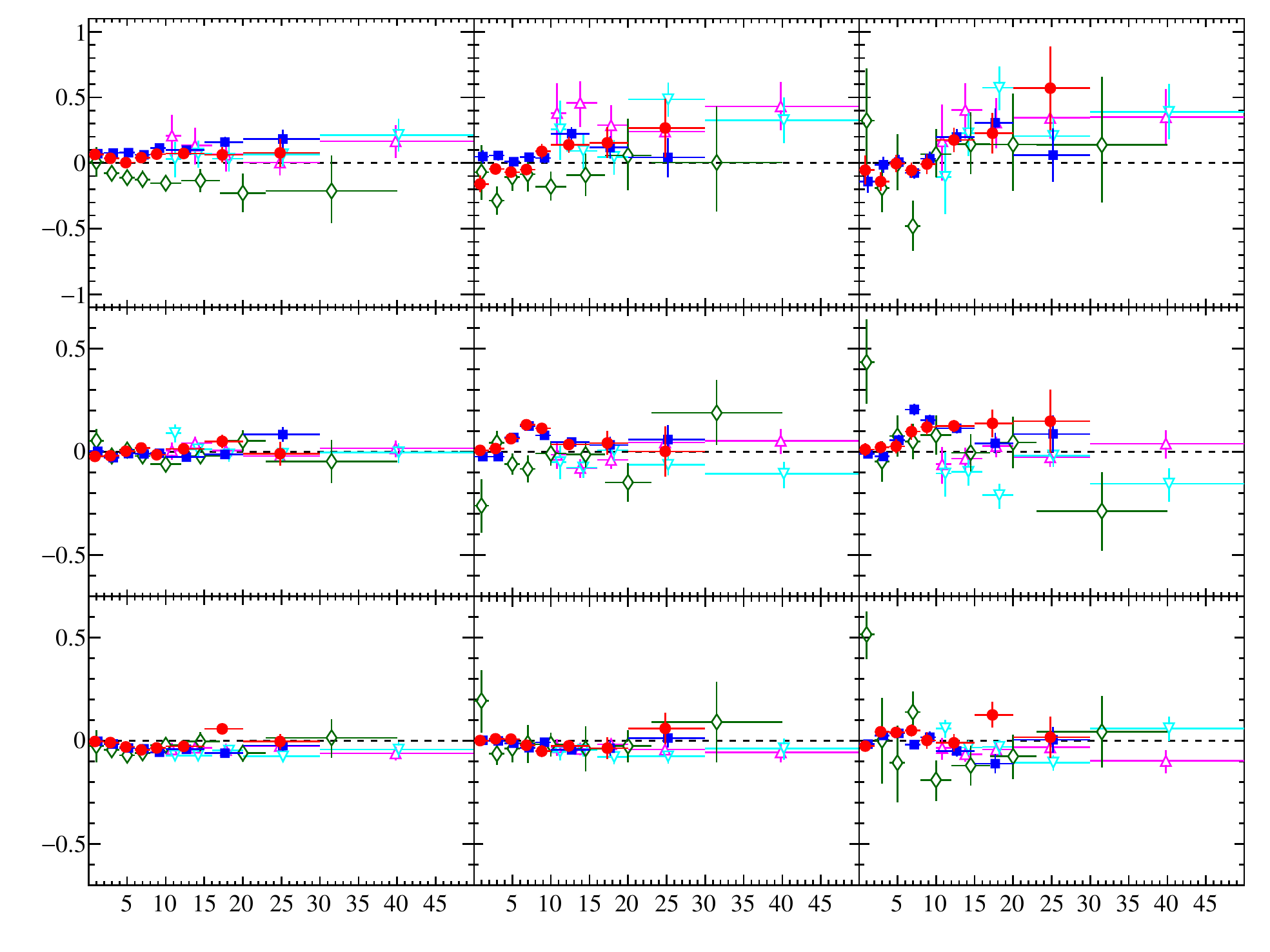}
    }
    \put ( 14, 88){{${\color{red}\bullet}$}                      {\tiny LHCb~$\sqrt{s}=7\tev~2.2<y<4.5$}   }
    \put ( 14, 84){{\tiny${\color{blue}\blacksquare}$}           {\tiny LHCb~$\sqrt{s}=8\tev~2.2<y<4.5$}   }
    \put (105, 84){{\small${\color[rgb]{0,0.6,0}\Diamond}$} {\tiny CDF~$\sqrt{s}=1.96\tev~\left|y\right|<0.6$} }
    \put ( 59, 88){{\tiny${\color[rgb]{1,0,1}\bigtriangleup}$} {\tiny CMS~$\sqrt{s}=7\tev~\left|y\right|<0.6$} }
    \put ( 59, 84){{\tiny${\color[rgb]{0,1,1}\bigtriangledown}$} {\tiny CMS~$\sqrt{s}=7\tev~0.6<\left|y\right|<1.2$} }
    \put (  0,  98) { \begin{sideways}$\uplambda_{\theta}$\end{sideways}}
    \put (  0,  60) { \begin{sideways}$\uplambda_{\theta\phi}$\end{sideways}}
    \put (  0,  25) { \begin{sideways}$\uplambda_{\phi}$\end{sideways}}
    \put ( 32,  0) { $\pty$}  \put ( 43, 0) { $\left[\!\gevc\right]$}
    \put ( 75,  0) { $\pty$}  \put ( 88, 0) { $\left[\!\gevc\right]$}
    \put (120,  0) { $\pty$}  \put (134, 0) { $\left[\!\gevc\right]$}
    \put ( 15,112) {HX frame}
    \put ( 44,112) {\ones}
    \put ( 90,112) {\twos}
    \put (135,112) {\threes}
  \end{picture}
  \caption { \small
    The~values of
    $\uplambda_{\theta}$\,(top),
    $\uplambda_{\theta\phi}$\,(middle) and
    $\uplambda_{\phi}$\,(bottom) parameters,
    measured in the~HX~frame 
    for \ones\,(left),
    \twos\,(center) and
    \threes\,(right)~~mesons.
    Results of the \lhcb analysis
    for the rapidity region \mbox{$2.2<\yy<4.5$}
    are shown with
    red solid  circles and blue solid squares for data collected
    in $\proton\proton$~collisions
    at~\mbox{$\sqrt{s}=7$} and~\mbox{$8\,\mathrm{TeV}$}, respectively. 
    The~results by the~CMS~collaboration~\cite{CMSpolar2013}
    obtained 
    in $\proton\proton$~collision at~\mbox{$\sqrt{s}=7\,\mathrm{TeV}$}    
    for rapidity regions~\mbox{$\left|\yy\right|<0.6$}
    and~\mbox{$0.6<\left|\yy\right|<1.2$}          
    are shown with                              
    magenta open upward triangles and           
    cyan open downward triangles, respectively.
    The~results obtained by the~CDF~collaboration~\cite{CDFpolar2012}
    in $\proton\antiproton$~collision at~\mbox{$\sqrt{s}=1.96\,\mathrm{TeV}$}    
    for rapidity region~\mbox{$\left|\yy\right|<0.6$}
    are shown with
    green open diamonds.
    Some data points are displaced from
    the~bin centers to improve visibility.    
    The~error bars indicate the~sum of
    the~statistical and systematic uncertainties added in quadrature.
  }\label{fig:comparison_HX}
\end{figure}

\begin{figure}[htb]
  \setlength{\unitlength}{1mm}
  \centering
  \begin{picture}(155,120)
    %
    \put(  0,  0){
      \includegraphics*[width=150mm,height=120mm,%
      ]{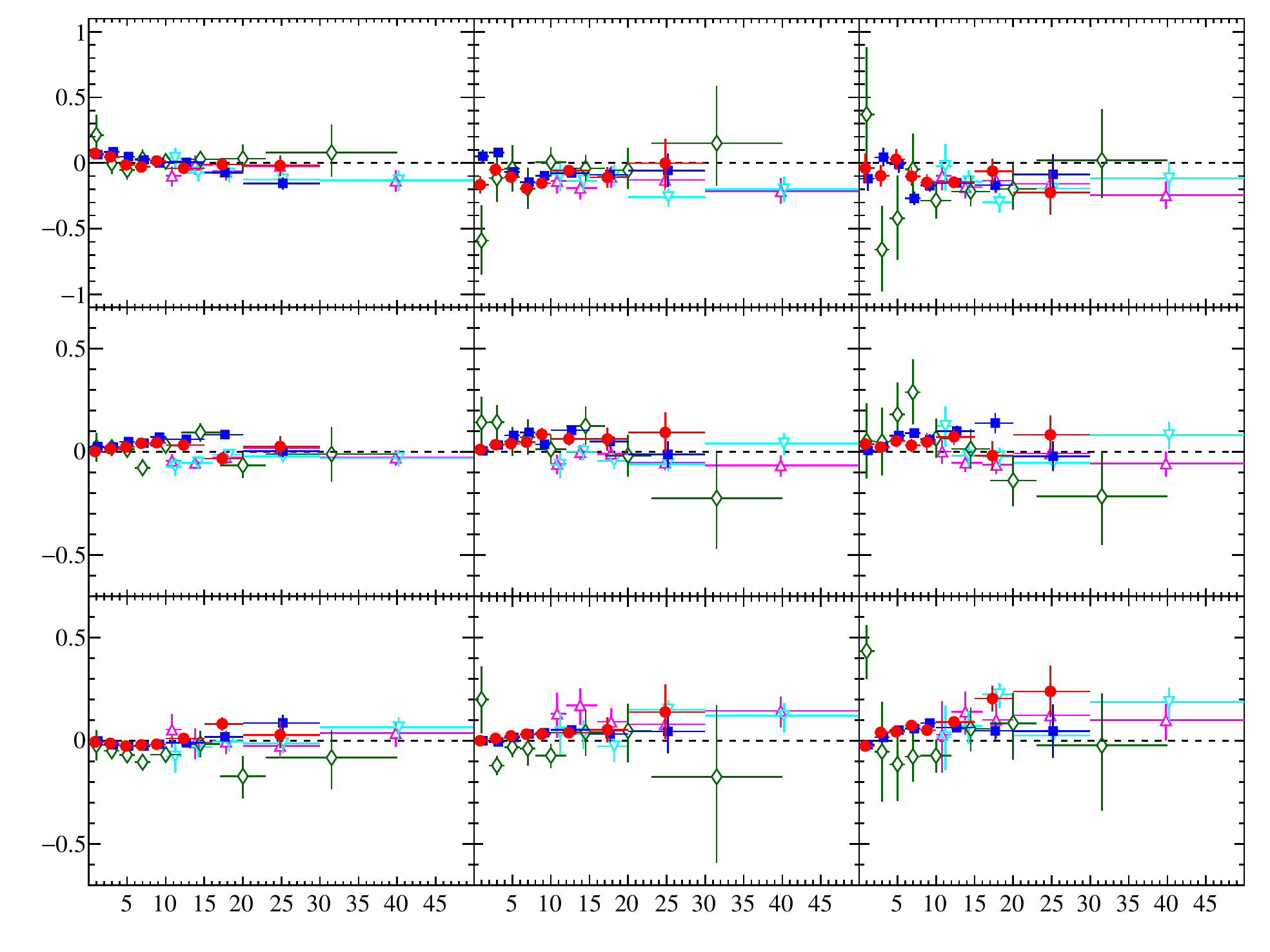}
    }
    \put ( 14, 88){{${\color{red}\bullet}$}                      {\tiny LHCb~$\sqrt{s}=7\tev~2.2<y<4.5$}   }
    \put ( 14, 84){{\tiny${\color{blue}\blacksquare}$}           {\tiny LHCb~$\sqrt{s}=8\tev~2.2<y<4.5$}   }
    \put (107, 84){{\small${\color[rgb]{0,0.6,0}\Diamond}$} {\tiny CDF~$\sqrt{s}=1.96\tev~\left|y\right|<0.6$} }
    \put ( 60, 88){{\tiny${\color[rgb]{1,0,1}\bigtriangleup}$} {\tiny CMS~$\sqrt{s}=7\tev~\left|y\right|<0.6$} }
    \put ( 60, 84){{\tiny${\color[rgb]{0,1,1}\bigtriangledown}$} {\tiny CMS~$\sqrt{s}=7\tev~0.6<\left|y\right|<1.2$} }
    \put (  0,  98) { \begin{sideways}$\uplambda_{\theta}$\end{sideways}}
    \put (  0,  60) { \begin{sideways}$\uplambda_{\theta\phi}$\end{sideways}}
    \put (  0,  25) { \begin{sideways}$\uplambda_{\phi}$\end{sideways}}
    \put ( 32,  0) { $\pty$}  \put ( 43, 0) { $\left[\!\gevc\right]$}
    \put ( 75,  0) { $\pty$}  \put ( 88, 0) { $\left[\!\gevc\right]$}
    \put (120,  0) { $\pty$}  \put (134, 0) { $\left[\!\gevc\right]$}
    \put ( 15,112) {CS frame}
    \put ( 44,112) {\ones}
    \put ( 90,112) {\twos}
    \put (135,112) {\threes}
  \end{picture}
  \caption { \small
    The~values of
    $\uplambda_{\theta}$\,(top),
    $\uplambda_{\theta\phi}$\,(middle) and
    $\uplambda_{\phi}$\,(bottom) parameters,
    measured in the~CS~frame 
    for \ones\,(left),
    \twos\,(center) and
    \threes\,(right)~~mesons.
    Results of the \lhcb analysis
    for the rapidity region \mbox{$2.2<\yy<4.5$}
    are shown with
    red solid  circles and blue solid squares for data collected
    in $\proton\proton$~collisions
    at~\mbox{$\sqrt{s}=7$} and~\mbox{$8\,\mathrm{TeV}$}, respectively. 
    The~results by the~CMS~collaboration~\cite{CMSpolar2013}
    obtained 
    in $\proton\proton$~collision at~\mbox{$\sqrt{s}=7\,\mathrm{TeV}$}    
    for rapidity regions~\mbox{$\left|\yy\right|<0.6$}
    and~\mbox{$0.6<\left|\yy\right|<1.2$}          
    are shown with                              
    magenta open upward triangles and           
    cyan open downward triangles, respectively.
    The~results obtained by the~CDF~collaboration~\cite{CDFpolar2012}
    in $\proton\antiproton$~collision at~\mbox{$\sqrt{s}=1.96\,\mathrm{TeV}$}    
    for rapidity region~\mbox{$\left|\yy\right|<0.6$}
    are shown with
    green open diamonds.
    Some data points are displaced from
    the~bin centers to improve visibility.    
    The~error bars indicate the~sum of
    the~statistical and systematic uncertainties added in quadrature.
  }\label{fig:comparison_CS}
\end{figure}

Figs.~\ref{fig:comparison_HX} and~\ref{fig:comparison_CS} show
a comparison of the \lhcb results~\cite{LHCb_YnS_polar_2017}
with results obtained
by the \cdf~\cite{CDFpolar2012} and \cms~\cite{CMSpolar2013}
collaborations
for the~HX  and CS~frames, respectively.
There is good agreement with \cms results for both frames,
and with \cdf for the CS frame.

The \lhcb collaboration continues to perform measurements
devoted to \bquark-hadron and quarkonium
physics~\cite{Monica2013,Manca2014,ZYang2014,Belyaev_Egorychev_2015}.
In particular,
the polarization studies performed by
\lhcb~\cite{LHCb_psi_polar,LHCb_psi2S_polar,LHCb_YnS_polar_2017}
allowed one to shed some new facts on production mechanism of heavy
quarkonium states, and to aggravate old ones.

\ack
I wish to thank A.V.~Efremov and O.V.~Teryaev
for their invitation to the DSPIN-17 workshop and
for their warm hospitality,
and also S.R.~Slabospitsky and O.V.~Teryaev for useful and
interesting discussions.

\medskip
\section*{References}
\medskip

\end{document}